\makeatletter
\@ifundefined{@parse@version@dash}{%
	\def\@parse@version#1{\@parse@version@0#1}
	\def\@parse@version@#1/#2/#3#4#5\@nil{%
		\@parse@version@dash#1-#2-#3#4\@nil}
	\def\@parse@version@dash#1-#2-#3#4#5\@nil{%
		\if\relax#2\relax\else#1\fi#2#3#4 }
}{}
\makeatother

\documentclass[aps,pre,twocolumn,groupedaddress,longbibliography]{revtex4-2}
\usepackage{amsfonts,amssymb,amsmath} 
\usepackage{color} 
\usepackage{graphicx} 
\usepackage{epstopdf,mathrsfs} 
\usepackage[colorlinks,linktocpage,linkcolor=blue]{hyperref}

\usepackage{xcolor}

\begin{document}

\title{Ergodic property of random diffusivity system with trapping events}
\author{Xudong Wang$^1$}
\author{Yao Chen$^2$}
\email{ychen@njau.edu.cn}
\affiliation{$^1$School of Mathematics and Statistics, Nanjing University of Science and Technology, Nanjing, 210094, P.R. China \\
$^2$College of Sciences, Nanjing Agricultural University, Nanjing, 210094, P.R. China}



\begin{abstract}
Brownian yet non-Gaussian phenomenon has recently been observed in many biological and active matter systems. The main idea of explaining this phenomenon is to introduce a random diffusivity for particles moving in inhomogeneous environment. This paper considers a Langevin system containing a random diffusivity and an $\alpha$-stable subordinator with $\alpha<1$. This model describes the particle's motion in complex media where both the long trapping events and random diffusivity exist.
We derive the general expressions of ensemble- and time-averaged mean-squared displacements which only contain the values of the inverse subordinator and diffusivity. Further taking specific time-dependent diffusivity, we obtain the analytic expressions of ergodicity breaking parameter and probability density function of the time-averaged mean-squared displacement. The results imply the nonergodicity of the random diffusivity model for any kind of diffusivity, including the critical case where the model presenting normal diffusion.

\end{abstract}

\maketitle

\section{Introduction}

In recent decades, it has been widely recognized that, beyond the classical Brownian motion, anomalous diffusion is a very general phenomenon in the natural world, which is characterized by the nonlinear evolution of ensemble-averaged mean-squared displacement (EAMSD) with respect to time, i.e.,
\begin{equation}
  \langle x^2(t)\rangle \simeq 2D_\beta t^\beta
\end{equation}
with $\beta\neq1$ \cite{HausKehr:1987,Bouchaud:1992,MetzlerKlafter:2000}. One of the common example of subdiffusion with $\beta<1$ is the continuous-time random walk (CTRW) with long trapping events characterized by power-law-distributed waiting times \cite{HeBurovMetzlerBarkai:2008,BurovJeonMetzlerBarkai:2011}. The simple models presenting superdiffusion with $\beta>1$ include L\'{e}vy flight with divergent second moment of jump length \cite{ShlesingerZaslavskyFrisch:1995,VahabiSchulzShokriMetzler:2013} and L\'{e}vy walk with heavy-tailed duration time of each running event \cite{TejedorMetzler:2010,MagdziarzMetzlerSzczotkaZebrowski:2012-1,ZaburdaevDenisovKlafter:2015,ChenWangDeng:2019}. There are still many anomalous diffusion processes, which present subdiffusion or superdiffusion depending on the specific value of system parameters, such as fractional Brownian motion \cite{MandelbrotNess:1968,DengBarkai:2009,MeerschaertSabzikar:2013,ChenWangDeng:2017}, scaled Brownian motion \cite{Safdari-etal:2015,ThielSokolov:2014,JeonChechkinMetzler:2014}, and heterogenous diffusion process \cite{CherstvyChechkinMetzler:2013,CherstvyMetzler:2013,CherstvyMetzler:2014,WangDengChen:2019}.
In addition, a large number of exotic and hybrid processes have been invented in recent years, such as the diffusivity can be exponentially increasing and decreasing in time or logarithmically increasing \cite{CherstvySafdariMetzler:2021}, or depending on both the position and time \cite{CherstvyMetzler:2015-2}, or a combined model of heterogenous diffusion process and fractional Brownian motion to describe the particle dynamics in complex systems with position-dependent diffusivity driven by fractional Gaussian noise \cite{Wang-etal:2020-2}.

A new class of diffusion dynamics has recently been observed in a large range of complex systems, which is named Brownian yet non-Gaussian process due to the coexisting phenomenon of linear EAMSD and non-Gaussian probability density function (PDF) \cite{WangAnthonyBaeGranick:2009,WangAnthonyBaeGranick:2009,ToyotaHeadSchmidtMizuno:2011,SilvaStuhrmannBetzKoenderink:2014,Bhattacharya-etal:2013,SamantaChakrabarti:2016}. The PDF of this new class of stochastic processes is characterized by an exponential distribution, rather than the Gaussian one.
The physical interpretation of the non-Gaussian PDF was given by a superstatistical approach \cite{Beck:2001,BeckCohen:2003,Beck:2006}, in other words, each particle undergoes the Brownian diffusion with its own diffusivity $D$ which does not change considerably in a short time less than the characteristic time of correlation in a system.
To explain the crossover from exponential distribution to Gaussian distribution of the Brownian non-Gaussian phenomena, Chubynsky and Slater proposed a diffusing diffusivity model with diffusivity undergoing a random walk \cite{ChubynskySlater:2014}, and Chechkin {\it et al.} established a minimal model with diffusing diffusivity under the framework of Langevin equation \cite{ChechkinSenoMetzlerSokolov:2017}.
To further describe the particle's stochastic motion in complex environments, the idea of random diffusivity has been applied to generalized Langevin equation \cite{SlezakMetzlerMagdziarz:2018}, generalized grey Brownian motion \cite{SposiniChechkinSenoPagniniMetzler:2018}, and fractional Brownian motion \cite{JainSebastian:2018,MackalaMagdziarz:2019}. Besides, the exponential tail is found to be universal for short-time dynamics of the CTRW by using large deviation theory  \cite{BarkaiBurov:2020,WangBarkaiBurov:2020}.

In this paper, we consider a coupled Langevin system with random diffusivity to describe the particle's motion in complex media where both the trapping events and random diffusivity exist. Instead of focusing on the PDF of the particle's trajectory, we
pay more attention to the (non)ergodic property of this coupled Langevin system by comparing the EAMSD and time-averaged mean-squared displacement (TAMSD) which is defined as \cite{MetzlerJeonCherstvyBarkai:2014,BurovJeonMetzlerBarkai:2011}
\begin{equation}\label{Def-TAMSD}
  \overline{\delta^2(\Delta)}=\frac{1}{T-\Delta}\int_{0}^{T-\Delta} [x(t+\Delta)-x(t)]^2 dt.
\end{equation}
It is usually assumed that the lag time $\Delta$ is much smaller than the total measurement time $T$ for obtaining a good statistical property. Based on the advance of single-particle tracking techniques, scientists often evaluate the recorded time series in terms of TAMSD to study the diffusion behavior of particles in living cell \cite{WeberSpakowitzTheriot:2010,GoldingCox:2006,BronsteinIsraelKeptenMaiTalBarkaiGarini:2009}.
A process is called ergodic if the TAMSD and EAMSD are equivalent, i.e., $\overline{\delta^2(\Delta)}=\langle x^2(\Delta)\rangle$ as $T\rightarrow\infty$, such as Brownian motion and (tempered) fractional Brownian motion \cite{Goychuk:2012,DengBarkai:2009,ChenWangDeng:2017}.
The ergodic property of a random diffusivity model has been partly discussed for some models, such as the model with local diffusivity fluctuating in time \cite{CherstvyMetzler:2016}, the one with a power-law correlated fractional Gaussian noise \cite{Wang-etal:2020}, and the one with superstatistical, uncorrelated or correlated diffusivity \cite{WangChen:2021}.

The scatter of the amplitude of TAMSD is also the main quantity to be studied in this paper, which is denoted by $\phi(\eta)$ with the dimensionless random variable $\eta$ defined as \cite{MetzlerJeonCherstvyBarkai:2014,BurovJeonMetzlerBarkai:2011}
\begin{equation}\label{Def-eta0}
  \eta:=\frac{\overline{\delta^2(\Delta)}}{\langle \overline{\delta^2(\Delta)}\rangle}.
\end{equation}
The dimensionless amplitude $\eta$ is the useful indicator to classify numerous anomalous diffusion processes by analyzing its statistics, which include the PDF of $\eta$ and the variance of $\eta$, i.e., ergodicity breaking (EB) parameter. These statistics for our random diffusivity model will be explicitly derived by taking specific diffusivity and subordinator in this paper.

The structure of this paper is as follows. In Sec. \ref{Sec2}, the random diffusivity model, together with the properties of the subordinator, are introduced. Based on this model, we derive the general expressions of the EAMSD and TAMSD in Sec. \ref{Sec3}. Further, by taking specific time-dependent diffusivity, we obtain the explicit expressions of the EAMSD and TAMSD in Sec. \ref{Sec4}, and the corresponding EB parameter together with the distribution of TAMSD in Sec. \ref{Sec5}. Some discussions are provided in Sec. \ref{Sec6}. For convenience, we put the algorithm of simulations and some mathematical details in Appendix.

\section{Random diffusivity model and subordinator}\label{Sec2}
We consider the following one-dimensional overdamped Langevin equation coupled with a subordinator:
\begin{equation}\label{DDmodel}
  \dot{x}(s)=\sqrt{2D(s)}\xi(s),  \quad \dot{t}(s)=\zeta(s),
\end{equation}
where $x(s)$ denotes the particle's trajectory over operational time $s$, $\xi(s)$ is a Gaussian white noise with null mean $\langle\xi(s)\rangle=0$ and the correlation function $\langle\xi(s_1)\xi(s_2)\rangle=\delta(s_1-s_2)$, and $\zeta(s)$ is a fully skewed $\alpha$-stable L\'{e}vy noise with $0<\alpha<1$ \cite{SchertzerLarchevequeDuanYanovskyLovejoy:2001} which is usually regarded as the formal derivative of the $\alpha$-stable subordinator $t(s)$ \cite{Applebaum:2009}. The dot over a one-variable function denotes the first derivative of this function with respect to this variable.
In this random diffusivity model, the diffusivity $D(s)$ can be a time-dependent random variable or a stochastic process \cite{WangChen:2021}. A deterministic diffusivity $D_0$ (i.e., $D(s)\equiv D_0$) in Eq. \eqref{DDmodel} leads to the classical Brownian motion coupled with a subordinator, which exhibits subdiffusion behavior and presents a stretched-Gaussian PDF \cite{MetzlerKlafter:2000}.

The Langevin equation can describe the particle's trajectory at any time, and it has the correspondence to another common physical model, CTRW.
The $\alpha$-stable subordinator $t(s)$ with $0<\alpha<1$ in this Langevin system corresponds to the CTRW where the particle gets immobilised for a trapping time drawn from the power-law-distributed waiting time with exponent $\alpha$. Note that the diffusivity is $D(s)$ in the first equation of Eq. \eqref{DDmodel}, rather than $D(t(s))$. It means that the diffusivity also remains unchanged when the particle gets immobilised for a trapping time.

The object process considered in model Eq. \eqref{DDmodel} is $x(t):=x(s(t))$, where $s(t)$ is named inverse subordinator being a one-to-one correspondence to the subordinator $t(s)$ \cite{KumarVellaisamy:2015,AlrawashdehKellyMeerschaertScheffler:2017}. The explicit mathematical definition of the inverse subordinator is
\begin{equation}
s(t)=\inf_{s>0}\{s:t(s)>t\}.
\end{equation}
The object process $x(t)$ here is also called subordinated process, compared with the original process $x(s)$. The subordinated process $x(t)$ is actually a recombination process of the original process $x(s)$ and the inverse subordinator $s(t)$. Therefore, the properties of the subordinated process $x(t)$ are determined by both $x(s)$ and $s(t)$.

The $\alpha$-stable subordinator $t(s)$ is a non-decreasing L\'{e}vy process with stationary and independent increments  \cite{Applebaum:2009}, and its characteristic function is
\begin{equation}
	\langle e^{-\lambda t(s)}\rangle=e^{-s\lambda^\alpha}.
\end{equation}
Based on the characteristic function and the property of the subordinator, the PDF $h(s,t)$ of the inverse $\alpha$-stable subordinator $s(t)$ can be obtained, which is usually expressed in Laplace domain ($t\rightarrow\lambda$) as \cite{BauleFriedrich:2005}
\begin{equation}\label{hslambda}
h(s,\lambda)=\int_0^\infty e^{-\lambda t} h(s, t)dt=\lambda^{\alpha-1}e^{-s\lambda^\alpha}.
\end{equation}
Denote the PDF of the subordinated process $x(t)$ as $p(x,t)$ and the one of the original process $x(s)$ as $p_0(x,s)$. Due to the independence between the original process $x(s)$ and the inverse subordinator $s(t)$, the PDF $p(x,t)$ can be written as \cite{Barkai:2001,BauleFriedrich:2005,ChenWangDeng:2018-2,ChenWangDeng:2019-2}
\begin{equation}\label{PDF_subor}
p(x,t)=\int_0^\infty  p_0(x,s)h(s,t) ds.
\end{equation}
Multiplying $x^n$ on both sides and performing the integral over $x$, we obtain the moments of the subordinated process in the expression of that of the original process:
\begin{equation}\label{hst1}
  \langle x^n(t)\rangle=\int_0^\infty  \langle x^n(s)\rangle h(s,t) ds,
\end{equation}
where $\langle x^n(s)\rangle$ represents the moments of the original process $x(s)$ on operational time $s$.
Similarly, the two-point joint PDF $p(x_2,t_2;x_1,t_1)$ of the subordinated process $x(t)$ can be obtained through the two-point joint PDF $p_0(x_2,s_2;x_1,s_1)$ of the original process $x(s)$ and the two-point joint PDF $h(s_2,t_2;s_1,t_1)$ of the inverse subordinator $s(t)$,
\begin{equation}\label{hst2}
\begin{split}
&p(x_2,t_2;x_1,t_1)\\
&~~=\int_0^\infty \int_0^\infty p_0(x_2,s_2;x_1,s_1)h(s_2,t_2;s_1,t_1)ds_2ds_1.
\end{split}
\end{equation}
Comparing with the explicit expression of $h(s,t)$ in Laplace domain in Eq. \eqref{hslambda}, the Laplace expression $(t_1\rightarrow\lambda_1,t_2\rightarrow\lambda_2)$ $h(s_2,\lambda_2;s_1,\lambda_1)$ is a bit more complicated \cite{BauleFriedrich:2005}, and not shown here. Fortunately, when evaluating the autocorrelation function of model Eq. \eqref{DDmodel}, the two-point joint PDF $h(s_2,t_2;s_1,t_1)$ can be reduced to the single point one as Eq. \eqref{ACF-xt} shows, i.e.,
\begin{equation}\label{hst2-1}
	\int_0^\infty h(s_2,t_2;s_1,t_1) ds_2=h(s_1,t_1),
\end{equation}
and the calculation gets simplified.

\section{EAMSD and TAMSD}\label{Sec3}
When the diffusivity $D(s)$ returns to a constant, the model Eq. \eqref{DDmodel} describes a Langevin system equivalent to the continuum limit of the CTRW with power-law-distributed waiting times, which has been discussed a lot \cite{KlafterSilbey:1980,Nelson:1999,BeckerMeerschaertScheffler:2004,Scalas:2006,VotAbadYuste:2017,AkimotoCherstvyMetzler:2018}. Therefore, we try to find out how the diffusivity $D(s)$ influences the diffusion behavior of the model Eq. \eqref{DDmodel}, and to establish the quantitative relation between them.

For a time-dependent diffusivity $D(s)$, we can directly perform integral on the first equation in Eq. \eqref{DDmodel} to obtain the expression of trajectory $x(s)$ and further take ensemble average to obtain the EAMSD of original process $x(s)$, i.e.,
\begin{equation}\label{EAMSDs}
  \begin{split}
    \langle x^2(s)\rangle
    &=2\int_0^sds_1'\int_0^sds_2'\left\langle \sqrt{D(s_1')D(s_2')}\xi(s_1')\xi(s_2')\right\rangle \\
    &=2\int_0^sds_1'\int_0^sds_2'\left\langle \sqrt{D(s_1')D(s_2')}\right\rangle \delta(s_1'-s_2') \\
    &=2\int_0^s  \langle D(s')\rangle ds',
  \end{split}
\end{equation}
where the independence between diffusivity $D(s)$ and white noise $\xi(s)$, together with the $\delta$-correlation function of $\xi(s)$ have been used in the second line of Eq. \eqref{EAMSDs}.

Substituting the $\langle x^2(s)\rangle$ in Eq. \eqref{EAMSDs} into Eq. \eqref{hst1} and exchanging the order of integration between $s'$ and $s$ yields
\begin{equation}\label{EAMSDt2}
  \begin{split}
    \langle x^2(t)\rangle
    &=2\int_0^\infty \int_0^s\langle D(s')\rangle h(s,t)ds'ds \\
    &=2\int_0^\infty \int_{s'}^\infty\langle D(s')\rangle h(s,t)dsds'.
  \end{split}
\end{equation}
Then performing Laplace transform ($t\rightarrow\lambda$) on both sides of Eq. \eqref{EAMSDt2}, with the help of the expression of $h(s,\lambda)$ in Eq. \eqref{hslambda}, we obtain
\begin{equation}\label{EAMSDt3}
  \begin{split}
    \mathcal{L}_{t\rightarrow\lambda}\langle x^2(t)\rangle
    &=\frac{2}{\lambda}\int_0^\infty \langle D(s')\rangle e^{-s'\lambda^\alpha} ds'  \\
    &=\frac{2}{\lambda}\mathcal{L}_{s\rightarrow\lambda^\alpha}\langle D(s)\rangle.
  \end{split}
\end{equation}
The Eq. \eqref{EAMSDt3} implies that the EAMSD only depends on the mean value of diffusivity $D(s)$. The subdiffusion behavior $\langle x^2(t)\rangle\propto t^\alpha$ is recovered when diffusivity is a time-independent random variable or the mean $\langle D(s)\rangle$ tends to a constant for long time. Otherwise, the diffusion behavior of the random diffusivity model is closely related to the trend of mean $\langle D(s)\rangle$. Due to the existence of subordinator, the effect of the diffusivity $D(s)$ on diffusion behavior works through the scaling factor $\lambda^\alpha$ rather than $\lambda$ in frequency domain, as Eq. \eqref{EAMSDt3} shows.

Now we turn to the TAMSD of the process described by Langevin equation \eqref{DDmodel}. As the definition of TAMSD in Eq. \eqref{Def-TAMSD} shows, the autocorrelation function of position $\langle x(t_1)x(t_2)\rangle$ is needed. Using the property of $\delta$-correlation of white noise $\xi(s)$, the autocorrelation function of $x(s)$ on operational time $s$ satisfies
\begin{equation}\label{ACF-xs}
  \langle x(s_1)x(s_2) \rangle= \langle x^2(s_1) \rangle
\end{equation}
for $s_1<s_2$. Then by use of the two-point PDF $h(s_2,t_2;s_1,t_1)$ of the inverse subordinator $s(t)$ and Eqs. \eqref{hst2} and \eqref{hst2-1}, we obtain
\begin{equation}\label{ACF-xt}
\begin{split}
    \langle x(t_1)x(t_2)\rangle
    &= \int_0^\infty\int_0^\infty \langle x(s_1)x(s_2)\rangle h(s_2,t_2;s_1,t_1)ds_1ds_2 \\
    &= \int_0^\infty \langle x^2(s_1)\rangle h(s_1,t_1)ds_1 \\
    &=\langle x^2(t_1)\rangle
\end{split}
\end{equation}
for $t_1\leq t_2$. For both the subordinated process $x(t)$ in Eq. \eqref{ACF-xt} and the original process $x(s)$ in Eq. \eqref{ACF-xs},
the equivalence between autocorrelation function and EAMSD is essentially resulted from the property of $\delta$-correlation of white noise $\xi(s)$ in model Eq. \eqref{DDmodel}.

Then we substitute Eq. \eqref{ACF-xt} into the definition of TAMSD in Eq. \eqref{Def-TAMSD}, and thus obtain the ensemble-averaged TAMSD
\begin{equation}\label{EATAMSD1}
  \begin{split}
    \langle\overline{\delta^2(\Delta)}\rangle
    &=\frac{1}{T-\Delta}\int_{0}^{T-\Delta} \langle x^2(t+\Delta)\rangle-\langle x^2(t)\rangle dt.
  \end{split}
\end{equation}
It is not convenient to perform Laplace transform for further calculations as Eq. \eqref{EAMSDt3}. Instead, considering that the EAMSD $\langle x^2(t)\rangle$ is a function of $t$, the integrand can be estimated by use of the priori condition $\Delta\ll t$, which implies
\begin{equation}\label{Appro}
  \langle x^2(t+\Delta)\rangle-\langle x^2(t)\rangle
  \simeq \Delta \frac{d}{dt}\langle x^2(t)\rangle.
\end{equation}
Note that this special procedure is only valid for the case where $\langle x^2(t+\Delta)\rangle$ and $\langle x^2(t)\rangle$ are separable. Otherwise, the explicit autocorrelation function of position $x(t)$ in the overdamped case, or that of velocity $v(t)$ in the underdamped case, is needed to evaluate the ensemble-averaged TAMSD.
Therefore, Eq. \eqref{EATAMSD1} yields the result
\begin{equation}\label{EATAMSD2}
  \langle\overline{\delta^2(\Delta)}\rangle
  \simeq \frac{\Delta}{T}\langle x^2(T)\rangle,
\end{equation}
which presents a normal diffusion with respect to lag time $\Delta$.
This phenomenon makes the TAMSD deviate from the EAMSD which is anomalous when $\alpha\gamma\neq1$, and thus implies the weak ergodicity breaking.
For the critical case $\alpha\gamma=1$, although both the EAMSD and ensemble-averaged TAMSD present the same normal diffusion behavior, the TAMSD is not self-averaged due to the scale-free property of subordinator with $\alpha<1$, i.e., $\overline{\delta^2(\Delta)}\neq\langle\overline{\delta^2(\Delta)}\rangle$ as $T\rightarrow\infty$. The critical case with $\alpha\gamma=1$ will also be discussed in Sec. \ref{Sec5}. Therefore, the random diffusivity model Eq. \eqref{DDmodel} is weakly nonergodic for any $\alpha<1$.

In general, the TAMSD is a stochastic process due to the randomness of the integrand $[x(t+\Delta)-x(t)]^2$ in Eq. \eqref{Def-TAMSD}. The ensemble average on TAMSD miss the information of randomness of TAMSD. Therefore, as a more detailed quantity, the scatter of TAMSD is a useful indicator to distinguish various anomalous diffusion processes.

For subdiffusive CTRW with power-law-distributed waiting time, it can be observed quite often for a wide range of $t$ that no jump event happens between time $t$ and $t+\Delta$, which leads to $[x(t+\Delta)-x(t)]^2=0$ for many different values of $t$. By contrast, the jump occurs at any small time interval with respect to the operational time $s$ for the random diffusivity model Eq. \eqref{DDmodel}.
A constant diffusivity $D$ of classical Brownian motion indicates that the magnitude of jumps at each step are of the same size, whereas the random diffusivity $D(s)$ resembles the inhomogeneous magnitude of different jumps.
From another point of view, the discrepant magnitude of jumps can be regarded as varying numbers of jumps with the same magnitude and the variance of each jump is a finite constant. In this sense, the number of jumps between time $t$ and $t+\Delta$ can be written as
\begin{equation}
  2\int_{s(t)}^{s(t+\Delta)} D(s')ds',
\end{equation}
which is random due to the randomness of both diffusivity $D(s)$ and inverse subordinator $s(t)$. Combining it with the definition of TAMSD in Eq. \eqref{Def-TAMSD}, we find that the TAMSD behaves as
\begin{equation}\label{Dist-TAMSD}
  \overline{\delta^2(\Delta)}\simeq C\int_0^{T-\Delta} \int_{s(t)}^{s(t+\Delta)} D(s')ds' dt,
\end{equation}
where $C$ is a constant independent of diffusivity $D$ and it will be determined by taking ensemble average on both sides later. We firstly consider the condition $\Delta\ll T$ and the long time limit of $T$. Similar to Eq. \eqref{Appro}, the inner integral can be approximated by
\begin{equation}
  \int_{s(t)}^{s(t+\Delta)} D(s')ds'
  \simeq \Delta \dot{s}(t)D(s(t))
\end{equation}
for $\Delta\ll t$.
So we further obtain
\begin{equation}\label{TAMSD1}
  \overline{\delta^2(\Delta)}\simeq C\Delta\int_0^{s(T)} D(s')ds'.
\end{equation}
Performing ensemble average on both sides of Eq. \eqref{TAMSD1} leads to
\begin{equation}
  \langle \overline{\delta^2(\Delta)}\rangle \simeq
  C\Delta \int_0^\infty\int_0^{s} D(s')ds' h(s,T)ds.
\end{equation}
Compared it with Eqs. \eqref{EAMSDt2} and \eqref{EATAMSD2}, we find that $C=2/T$. Therefore, the asymptotic behavior of TAMSD is
\begin{equation}\label{TAMSD2}
  \overline{\delta^2(\Delta)}\simeq \frac{2\Delta}{T}\int_0^{s(T)} D(s')ds'.
\end{equation}

There are two random factors in Eq. \eqref{TAMSD2}, the inverse subordinator $s(T)$ and diffusivity $D(s')$. The $s(T)$ here can be regarded as the number of jumps happening in physical time $(0,T)$ for a particle.
If the diffusivity returns to the constant $D$, then $\overline{\delta^2(\Delta)}\simeq \frac{2D\Delta s(T)}{T}$, consistent to the result in Ref. \cite{HeBurovMetzlerBarkai:2008}, where the number of jumps is denoted by $N$. On the other hand, if the subordinator vanishes, i.e., $s(T)=T$, then $\overline{\delta^2(\Delta)}\simeq \frac{2\Delta}{T}\int_0^{T} D(s')ds'$ proportional to the time average of diffusivity $D(s)$, which is consistent to the result in Ref. \cite{WangChen:2021}.

\section{Time-dependent diffusivity}\label{Sec4}
From the discussions above, we find that the EAMSD and ensemble-averaged TAMSD are only determined by the mean diffusivity $\langle D(s)\rangle$. For convenience, let us assume that the diffusivity has the asymptotic behavior
\begin{equation}\label{MeanD}
  \langle D(s) \rangle\simeq \gamma s^{\gamma-1}\,(\gamma>0)
\end{equation}
for long time. In fact, the $D(s)$ in random diffusivity model can be a random variable or a stochastic process, and the $D(s)$ at different times can be correlated or uncorrelated \cite{WangChen:2021}. Here, it does not matter which type the diffusivity $D(s)$ is, it matters how the mean diffusivity $\langle D(s) \rangle$ behaves at long time.

\begin{figure}
\begin{minipage}{0.35\linewidth}
  \centerline{\includegraphics[scale=0.29]{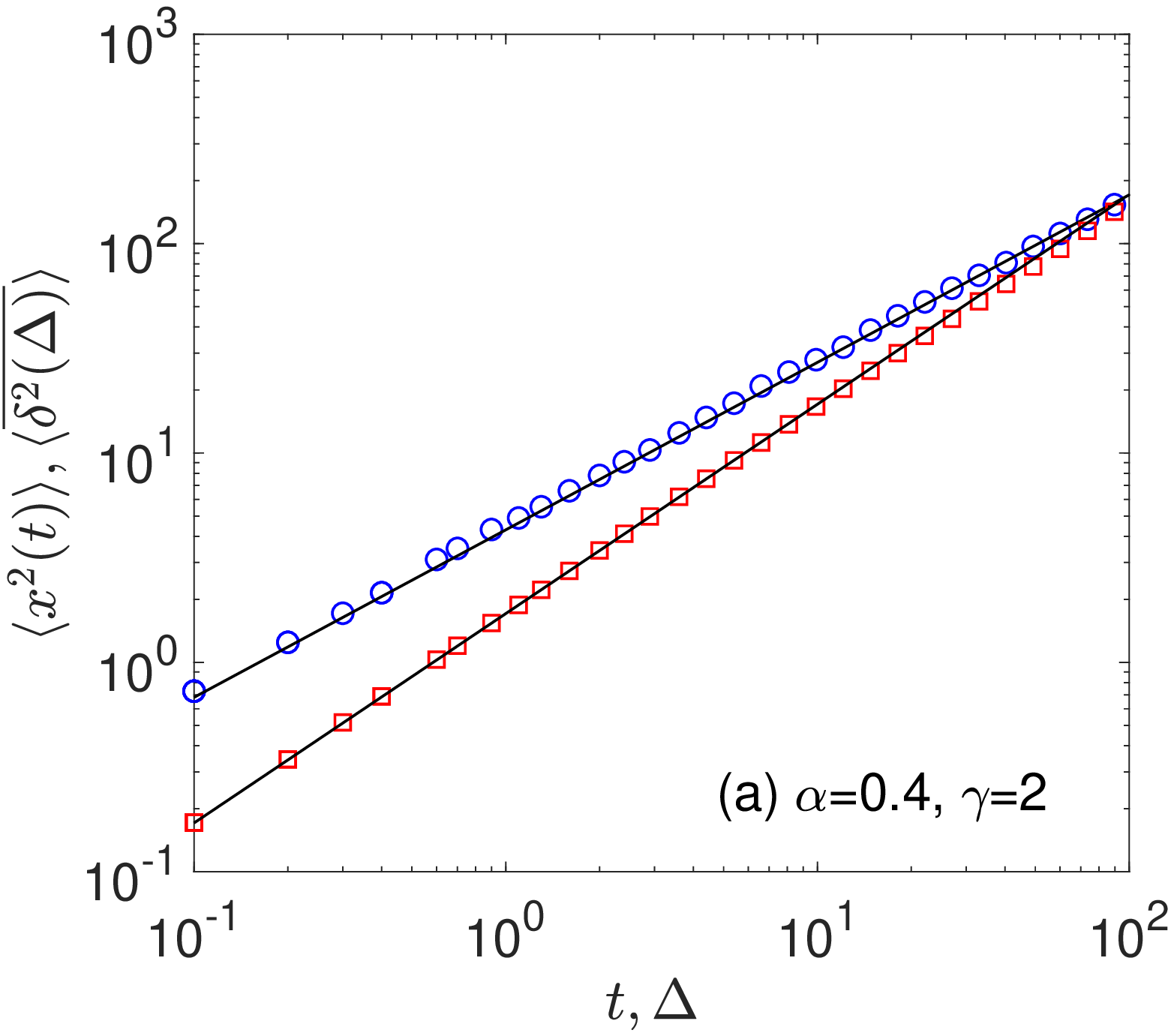}}
  \centerline{}
\end{minipage}
\hspace{1.43cm}
\begin{minipage}{0.35\linewidth}
  \centerline{\includegraphics[scale=0.29]{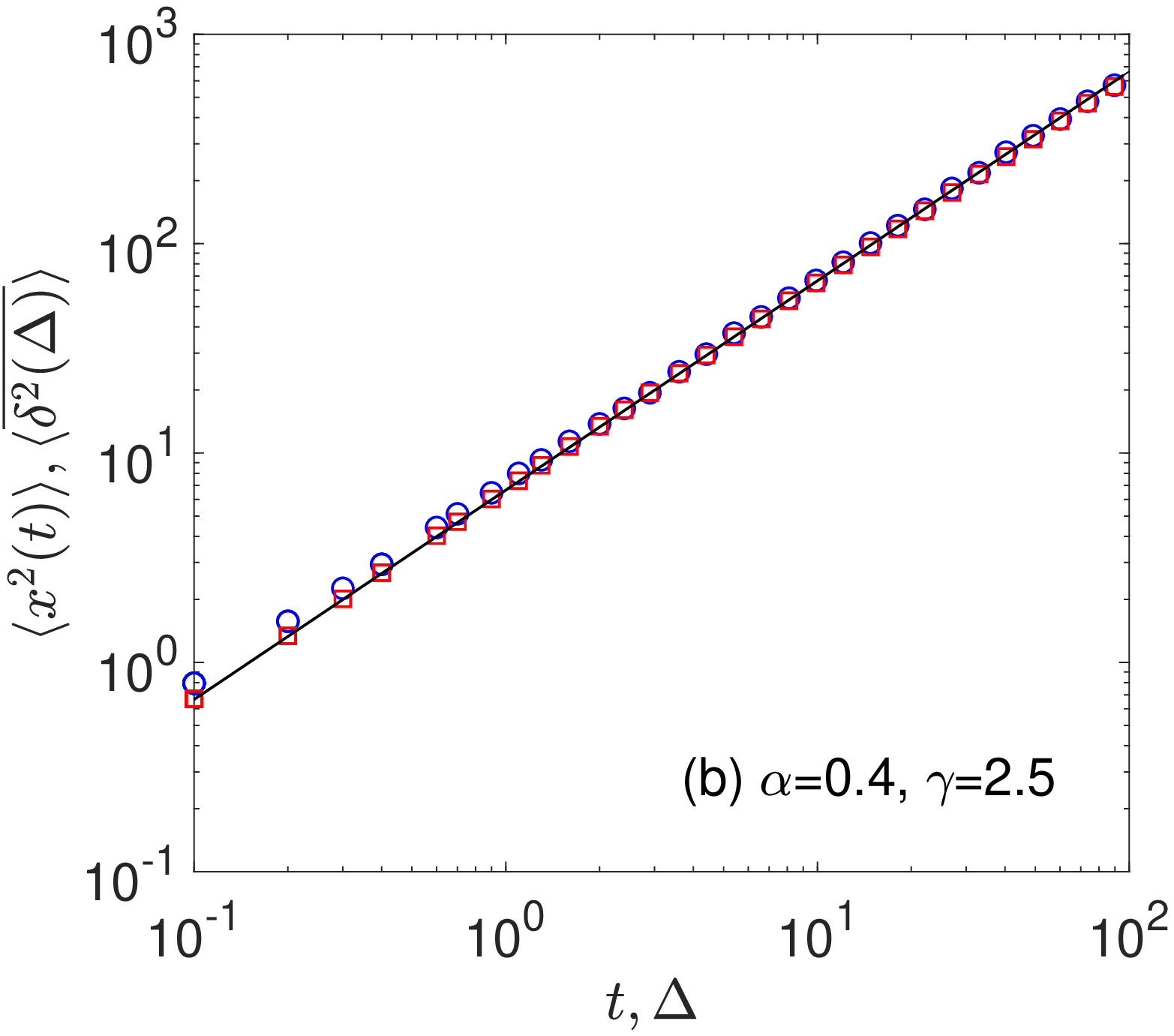}}
  \centerline{}
\end{minipage}
\begin{minipage}{0.35\linewidth}
  \centerline{\includegraphics[scale=0.29]{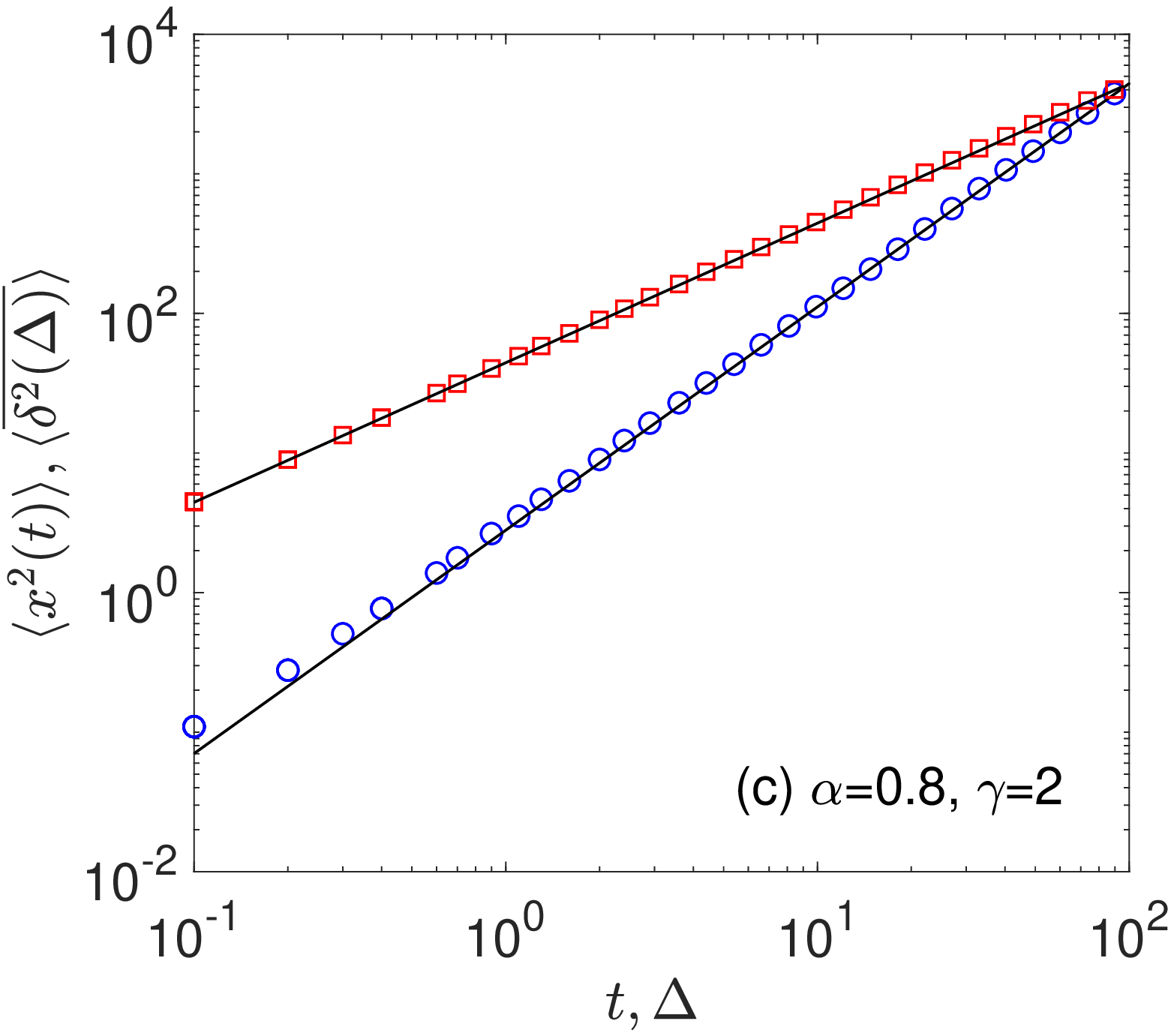}}
  \centerline{}
\end{minipage}
\hspace{1.43cm}
\begin{minipage}{0.35\linewidth}
  \centerline{\includegraphics[scale=0.29]{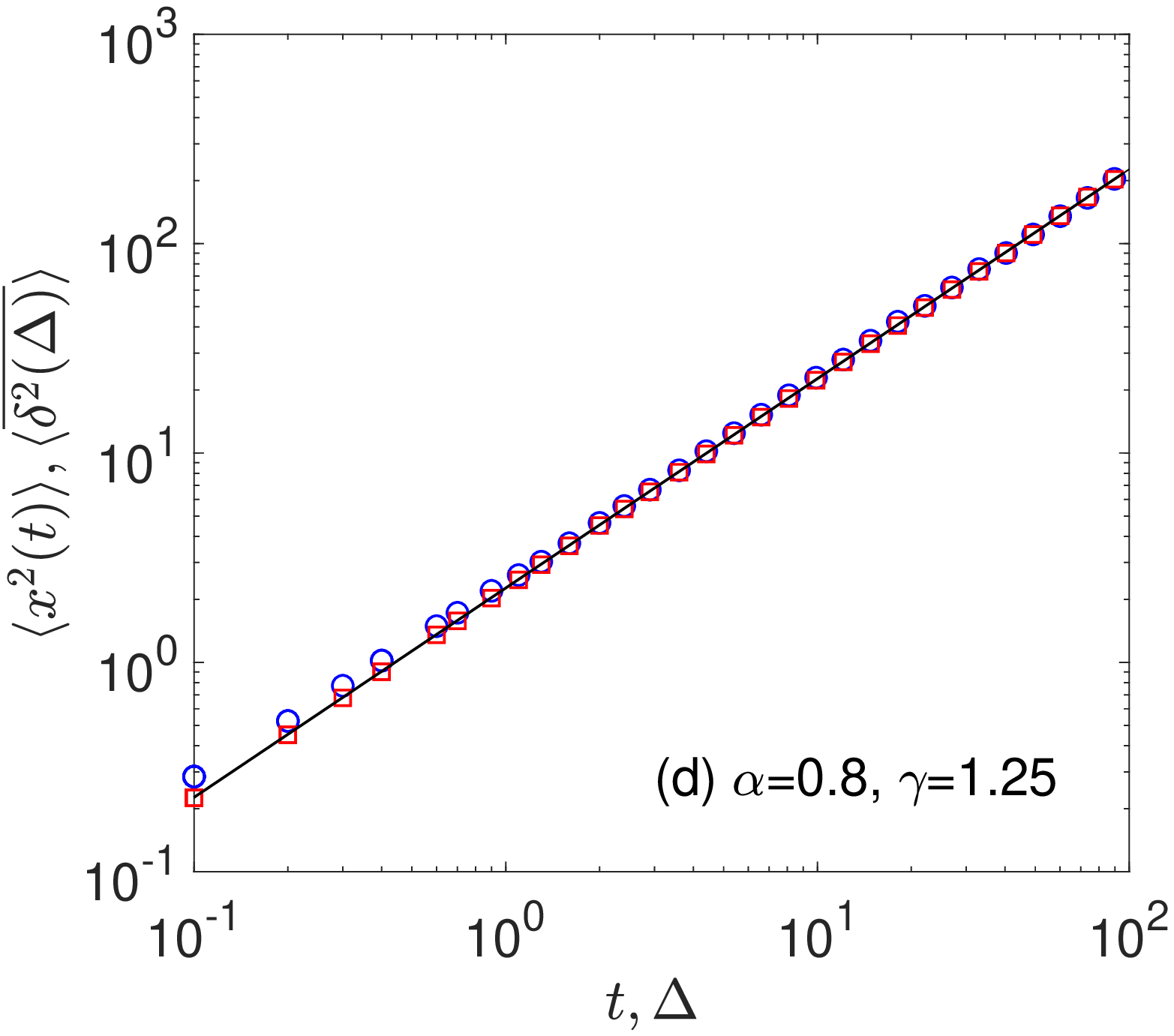}}
  \centerline{}
\end{minipage}
\caption{(Color online) EAMSD $\langle x^2(t)\rangle$ and ensemble-averaged TAMSD $\langle\overline{\delta^2(\Delta)}\rangle$ for random diffusivity model Eq. \eqref{DDmodel}. The theoretical results for $\langle x^2(t)\rangle$ in Eq. \eqref{EAMSD-2} and $\langle\overline{\delta^2(\Delta)}\rangle$ in Eq. \eqref{EATAMSD} are shown by black solid lines. The blue circles and red squares denote the simulated EAMSD and ensemble-averaged TAMSD, respectively. They fit to the theoretical lines perfectly in four pictures with different $\alpha$ and $\gamma$.
Other parameters: the measurement time is $T=10^2$, and the number of trajectories used for ensemble is $10^4$.}\label{fig1}
\end{figure}

Now we calculate the EAMSD and ensemble averaged TAMSD by use of the mean diffusivity $\langle D(s) \rangle$.
Substituting Eq. \eqref{MeanD} into Eq. \eqref{EAMSDt3}, we get
\begin{equation}\label{EAMSD-1}
   \mathcal{L}_{t\rightarrow\lambda}\langle x^2(t)\rangle
   =2\Gamma(\gamma+1)\lambda^{-\alpha\gamma-1}.
\end{equation}
Performing the inverse Laplace transform yields
\begin{equation}\label{EAMSD-2}
  \begin{split}
    \langle x^2(t)\rangle\simeq
    \frac{2\Gamma(\gamma+1)}{\Gamma(\alpha\gamma+1)}t^{\alpha\gamma}.
  \end{split}
\end{equation}
While the inverse subordinator with $\alpha<1$ slows down the diffusion behavior of particles, the time-dependent diffusivity $D(t)$ with $\gamma>0$ can both suppress and enhance the diffusion. The random diffusivity model presents subdiffusion when $\alpha\gamma<1$, and superdiffusion when $\alpha\gamma>1$. The normal diffusion $\langle x^2(t)\rangle=2\Gamma(\gamma+1)t$ is recovered at $\alpha\gamma=1$.
Similarly, the ensemble-averaged TAMSD in Eq. \eqref{EATAMSD2} is
\begin{equation}\label{EATAMSD}
  \langle\overline{\delta^2(\Delta)}\rangle
  \simeq
  \frac{2\Gamma(\gamma+1)}{\Gamma(\alpha\gamma+1)}\frac{\Delta}{T^{1-\alpha\gamma}},
\end{equation}
being normal for any $\alpha$ and $\gamma$.
The ratio between ensemble-averaged TAMSD and EAMSD is
\begin{equation}
  \frac{\langle\overline{\delta^2(\Delta)}\rangle}{\langle x^2(\Delta)\rangle}\simeq \left(\frac{\Delta}{T}\right)^{1-\alpha\gamma},
\end{equation}
which is not equal to $1$ and implies the weak ergodicity breaking if $\alpha\gamma\neq1$.
When $\alpha\gamma=1$, it holds that
\begin{equation}
  \langle x^2(\Delta)\rangle\simeq
  \langle\overline{\delta^2(\Delta)}\rangle
  \simeq 2\Gamma(\gamma+1)\Delta
\end{equation}
for long time. But the TAMSD cannot converge to its ensemble average, which can be proved in the next section. Therefore, the random diffusivity model Eq. \eqref{DDmodel} is weakly nonergodic for any $\alpha<1$ and $\gamma>0$.
We simulate the EAMSD and ensemble-averaged TAMSD with four kinds of parameters $(\alpha,\gamma)$ in Fig. \ref{fig1}. The EAMSD can present both subdiffusion, superdiffusion and normal diffusion, while the ensemble-averaged TAMSD is normal for any parameter.
The simulation results fit to the theoretical results perfectly.

\section{EB parameter and PDF of TAMSD}\label{Sec5}

For ergodic Brownian motion, its TAMSD converges to a constant for long time. However, the TAMSDs of many anomalous diffusion processes are random variables and present pronounced trajectory-to-trajectory variations, such as L\'{e}vy walk \cite{FroembergBarkai:2013,FroembergBarkai:2013-2,GodecMetzler:2013}, L\'{e}vy flight \cite{VahabiSchulzShokriMetzler:2013,FroembergBarkai:2013-2,BurneckiWeron:2010}, quenched models \cite{Massignan-etal:2014,MiyaguchiAkimoto:2011}, heterogeneous diffusion processes \cite{CherstvyChechkinMetzler:2013,CherstvyMetzler:2014,WangDengChen:2019,LeibovichBarkai:2019} and so on. The stochasticity of TAMSD can be measured by the scatter of the dimensionless random variable $\eta$, which in our model is
\begin{equation}\label{Def-eta}
  \eta\simeq\frac{2\int_0^{s(T)} D(s')ds'}{\langle x^2(T)\rangle}
\end{equation}
for large $T$. This result is obtained by substituting Eqs. \eqref{EATAMSD2} and \eqref{TAMSD2} into Eq. \eqref{Def-eta0}. The mean of the dimensionless random variable $\eta$ is independent of lag time $\Delta$ and satisfies $\langle\eta\rangle=1$. The result in Eq \eqref{Def-eta} is universal for the random diffusivity model Eq. \eqref{DDmodel} with any kind of subordinator and diffusivity.

It holds that $\phi(\eta)=\delta(\eta-1)$ for an ergodic process, while a nonergodic process shows a broad distribution of $\eta$. A measure of the scatter of TAMSD is the variance of dimensionless random variable $\eta$, which is also named EB parameter:
\begin{equation}\label{Def-EB}
    \textrm{EB}=\langle \eta^2\rangle-1.
\end{equation}
The EB parameter of the ergodic Brownian process scales as $\Delta/T$, and tends to zero as $T\rightarrow\infty$. In contrast, the EB parameter converges to a nonzero constant for nonergodic process.
Here for the random diffusivity model Eq. \eqref{DDmodel} with $\eta$ in Eq. \eqref{Def-eta}, it holds that
\begin{equation}
  \langle \eta^2\rangle = \frac{4I(T)}{\langle x^2(T)\rangle^2},
\end{equation}
where
\begin{equation}\label{IT}
  I(T)=\int_0^\infty \int_0^s\int_0^s \langle D(s'_1)D(s'_2)\rangle ds'_1ds'_2 h(s,T)ds.
\end{equation}

\begin{figure}
  \centering
  \includegraphics[scale=0.5]{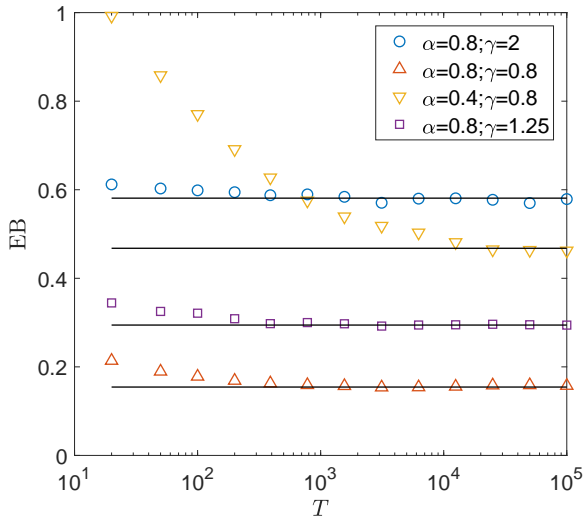}\\
  \caption{(Color online) EB parameters with four kinds of parameters ($\alpha,\gamma$).
  The markers (circle, square, positive and inverted triangles) denote the simulation results, while the four black solid lines represent the theoretical results in Eq. \eqref{EB1}.
  The total measurement time is $T=10^5$ and the lag time is $\Delta=1$. The number of trajectories used for ensemble is $10^4$. The simulation markers tend to the corresponding theoretical lines as $T\rightarrow\infty$.}\label{fig2}
\end{figure}

For convenience, we only consider that the diffusivity is uncorrelated at different times, i.e.,
\begin{equation}\label{DDcorr}
  \langle D(s_1)D(s_2)\rangle=\langle D(s_1)\rangle \langle D(s_2)\rangle
\end{equation}
for $s_1\neq s_2$. In this case, the EB parameter can be explicitly obtained as
\begin{equation}\label{EB1}
  \textrm{EB}\simeq\frac{\Gamma(1+2\gamma)\Gamma^2(1+\alpha\gamma)}
  {\Gamma(1+2\alpha\gamma)\Gamma^2(1+\gamma)}-1,
\end{equation}
where the detailed calculations are presented in Appendix \ref{App1}.
The simulations of the EB parameters are shown in Fig. \ref{fig2}, where four kinds of $\alpha$ and $\gamma$ are chosen. Whether $\alpha\gamma$ is equal to $1$ or not, the EB parameter converges to a constant at large $T$, consistent to the theoretical result in Eq. \eqref{EB1}.

Only when $\alpha=\gamma=1$, the EB parameter in Eq. \eqref{EB1} tends to zero and the random diffusivity process in Eq. \eqref{DDmodel} recovers the ergodic process. More precisely, $\alpha=1$ means that the power-law-distributed waiting times return to the exponential distribution, i.e., the long trapping events vanish, and $\gamma=1$ implies that the mean diffusivity keeps being a constant. In the case of $\alpha=\gamma=1$, the assumption of diffusivity being uncorrelated at different times makes the model converge to the classical Brownian motion. More discussions about uncorrelated and correlated diffusivity can be found in Ref. \cite{WangChen:2021}.

\begin{table}
\center
\caption{The EAMSD, ensemble-averaged TAMSD, and EB parameter as $T\rightarrow\infty$ for three specific nonergodic cases.}\label{table1}
\scalebox{1.20}{
\begin{tabular}{cccc}
  \hline
  Cases  & $\langle x^2(t)\rangle$ & $\langle\overline{\delta^2(\Delta)}\rangle$ & EB \\
\hline
  $\alpha=1,\gamma\neq1$ & $2t^\gamma$ & $\frac{2\Delta}{T^{1-\gamma}}$ & $0$ \\[5pt]

  $\alpha\neq1,\gamma=1$ & $\frac{2}{\Gamma(\alpha+1)}t^{\alpha}$ & $\frac{2}{\Gamma(\alpha+1)}\frac{\Delta}{T^{1-\alpha}}$ & $\frac{2\Gamma^2(1+\alpha)}{\Gamma(1+2\alpha)}-1$ \\[5pt]

  $\alpha\gamma=1$ & $2\Gamma(\gamma+1)t$ & $2\Gamma(\gamma+1)\Delta$ & $\frac{\Gamma(1+2\gamma)}{2\Gamma^2(1+\gamma)}-1$ \\[5pt]
  \hline
\end{tabular}}
\end{table}

On the contrast, the random diffusivity model is nonergodic once one of $\alpha$ and $\gamma$ is not equal to 1. There are three specific nonergodic cases, which are listed in Table \ref{table1}. The first nonergodic case is $\alpha=1,\gamma\neq1$, where the EAMSD is anomalous but the ensemble-averaged TAMSD is normal with respect to $\Delta$. However, it is worth noting that the EB parameter tends to zero as the measurement time $T\rightarrow\infty$. It means that the TAMSD converges to a constant for long time. In other words, the TAMSD has the property of self-averaging, i.e., $\overline{\delta^2(\Delta)}\simeq\langle\overline{\delta^2(\Delta)}\rangle$ as $T\rightarrow\infty$. This property is resulted from $\alpha=1$ and the assumption of uncorrelated diffusivity.

The second nonergodic case is $\alpha\neq1,\gamma=1$, where the EAMSD presents subdiffusion but the ensemble-averaged TAMSD is normal with respect to $\Delta$. In contrast to the first case, the EB parameter tends to a nonzero constant here, which implies a more significant nonergodicity than the first case. Moreover, the EAMSD, the ensemble-averaged TAMSD, and the EB parameter are consistent to the subdiffusive CTRW model in Ref. \cite{HeBurovMetzlerBarkai:2008} with the same exponent $\alpha<1$ of power-law-distributed waiting time. The essential reason is that the uncorrelated random diffusivity acts as a deterministic one when we perform the time averaging. For $\gamma=1$, the random diffusivity model actually converges to the one with constant diffusivity in the sense of evaluating the TAMSD.

The last nonergodic case is $\alpha\gamma=1$ ($\alpha<1,\gamma>1$), where both the EAMSD and ensemble-averaged TAMSD exhibit the normal diffusion with $\alpha\gamma=1$ as mentioned below Eq. \eqref{EATAMSD2}. In the prior condition of $\alpha<1$, however, the EB parameter is
\begin{equation}\label{EB2}
  \textrm{EB}\simeq\frac{\Gamma(1+2\gamma)}
  {2\Gamma^2(1+\gamma)}-1,
\end{equation}
which increases monotonously with respect to $\gamma$ when $\gamma>0$. Therefore, it is positive for $\gamma>1$ and implies the nonergodic behavior for the case $\alpha\gamma=1$ with $\alpha<1$.

Another quantity, the PDF of TAMSD $\phi(\eta)$, can measure the scatter of TAMSD directly. For an ergodic process, such as Brownian motion, the TAMSD is self-averaged and converge to a deterministic variable, which is embodied by the EB parameter tending to zero and the PDF $\phi(\eta)$ converging to $\delta(\eta-1)$.
By contrast, for the subdiffusive CTRW with the power-law-distributed waiting times characterizing the long trapping events, the corresponding limiting distribution $\phi(\eta)$ as measurement time $T\rightarrow\infty$  is the Mittag-Leffler distribution:
\begin{equation}\label{MLD}
  \lim_{T\rightarrow\infty}\phi(\eta)= \frac{ {\Gamma^{1/\alpha}(1+\alpha)}}{\alpha\eta^{1+1/\alpha}}L_\alpha
  \left[\frac{\Gamma^{1/\alpha}(1+\alpha)}{\eta^{1/\alpha}}\right],
\end{equation}
which has been discussed in many references \cite{HeBurovMetzlerBarkai:2008,WangDengChen:2019,LeibovichBarkai:2019,AghionKesslerBarkai:2019}.
Here, $\alpha$ is the exponent of power-law distribution, and $L_\alpha(1)$ is the one-sided L\'{e}vy stable distribution whose Laplace pair is $\exp(-\lambda^\alpha)$ \cite{Feller:1971,Barkai:2001}.
For the $\phi(\eta)$ in Eq. \eqref{MLD}, the corresponding EB parameter is \cite{HeBurovMetzlerBarkai:2008}
\begin{equation}
  \textrm{EB} = \frac{2\Gamma^2(1+\alpha)}{\Gamma(1+2\alpha)}-1,
\end{equation}
the same as the second nonergodic case in Table \ref{table1}.

\begin{figure}
\begin{minipage}{0.35\linewidth}
  \centerline{\includegraphics[scale=0.3]{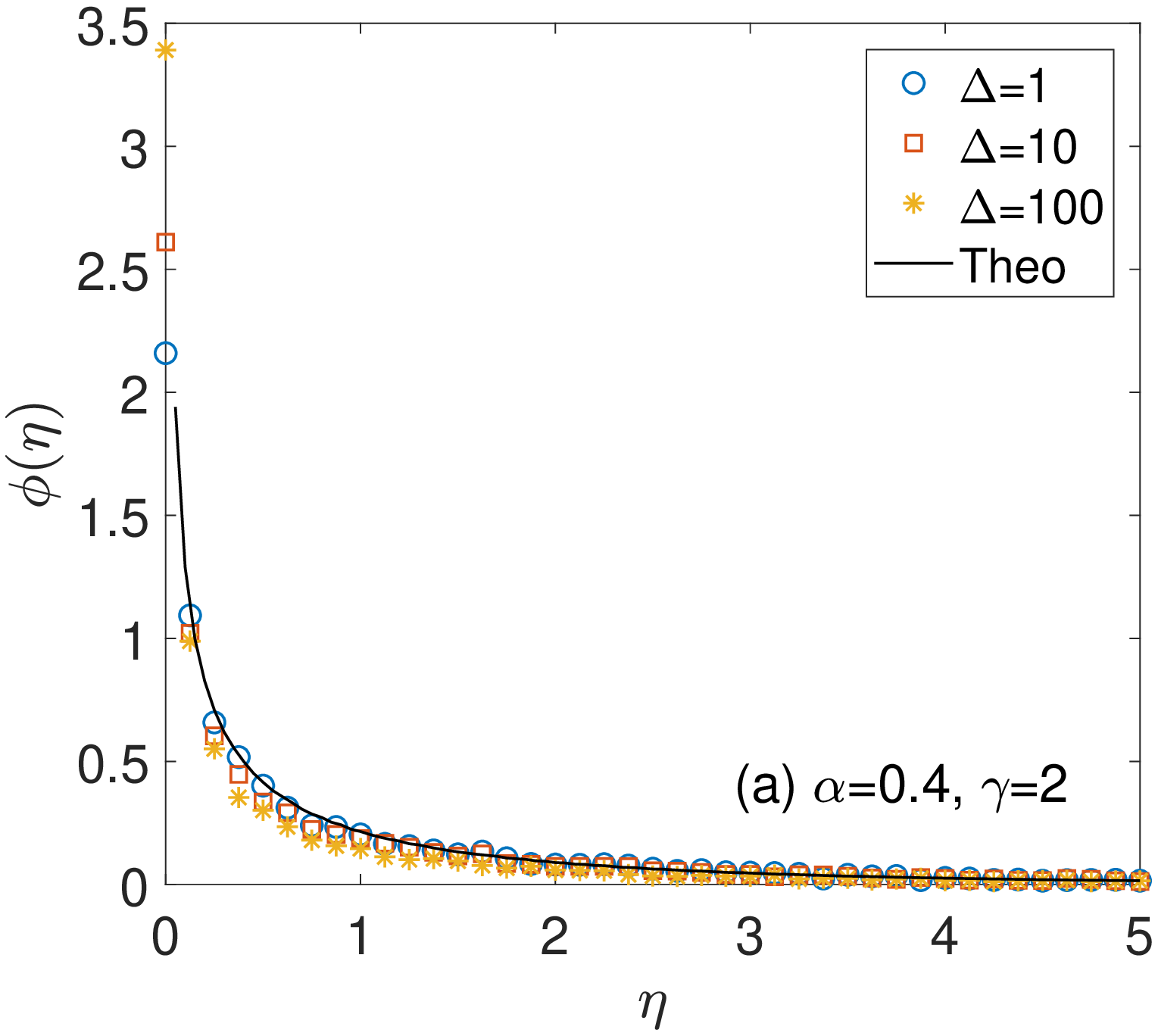}}
  \centerline{}
\end{minipage}
\hspace{1.4cm}
\begin{minipage}{0.35\linewidth}
  \centerline{\includegraphics[scale=0.3]{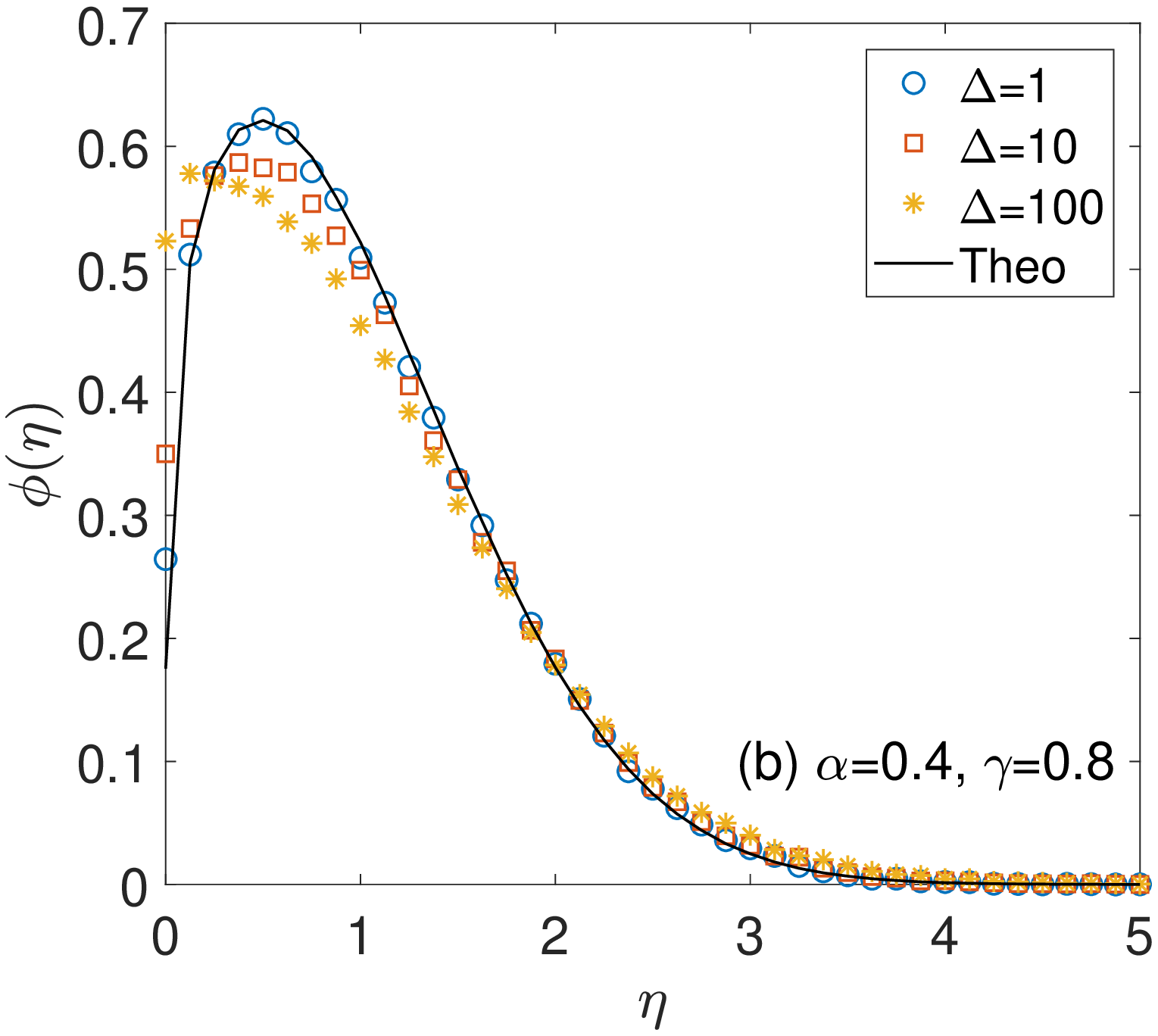}}
  \centerline{}
\end{minipage}
\begin{minipage}{0.35\linewidth}
  \centerline{\includegraphics[scale=0.3]{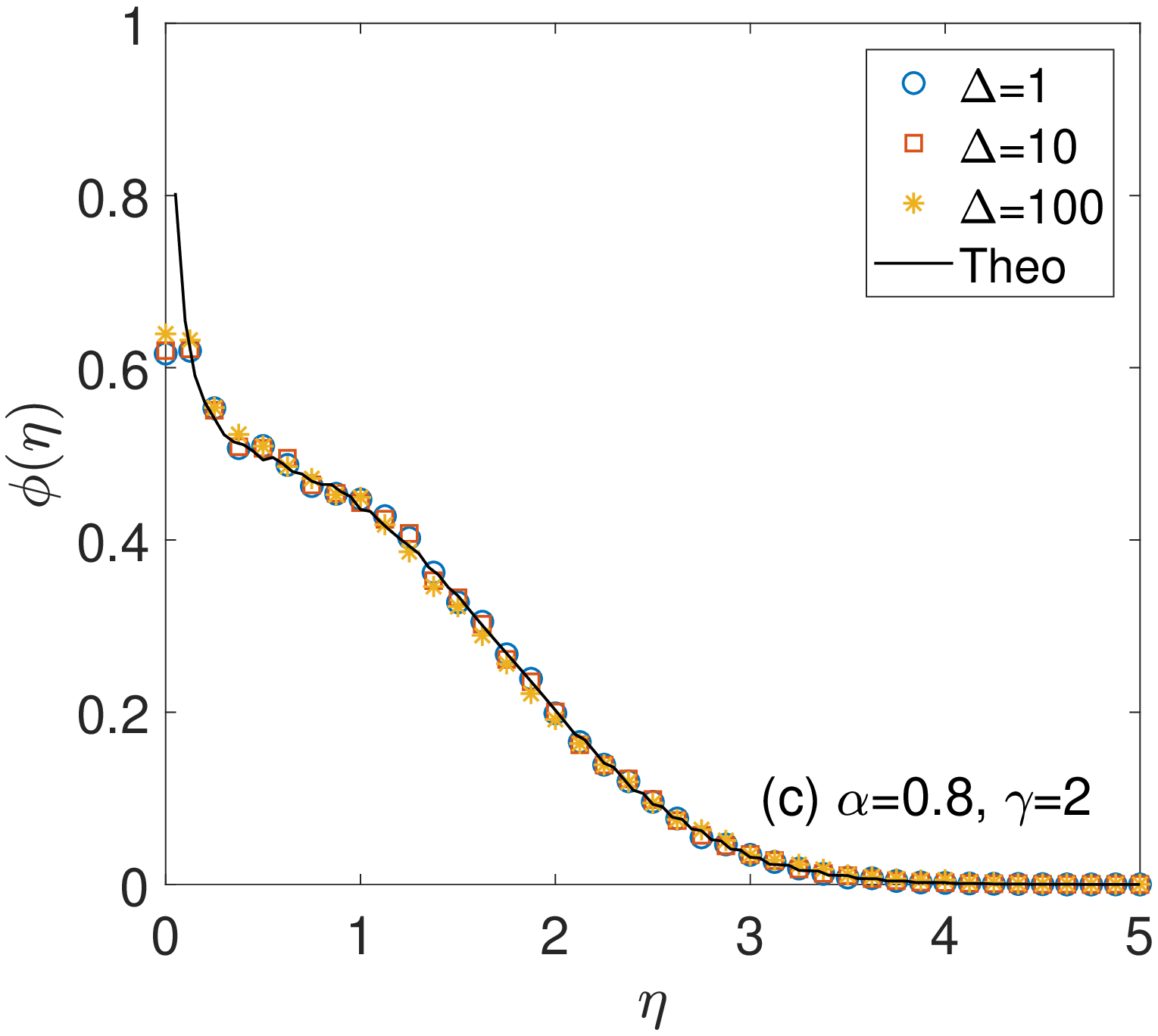}}
  \centerline{}
\end{minipage}
\hspace{1.4cm}
\begin{minipage}{0.35\linewidth}
  \centerline{\includegraphics[scale=0.3]{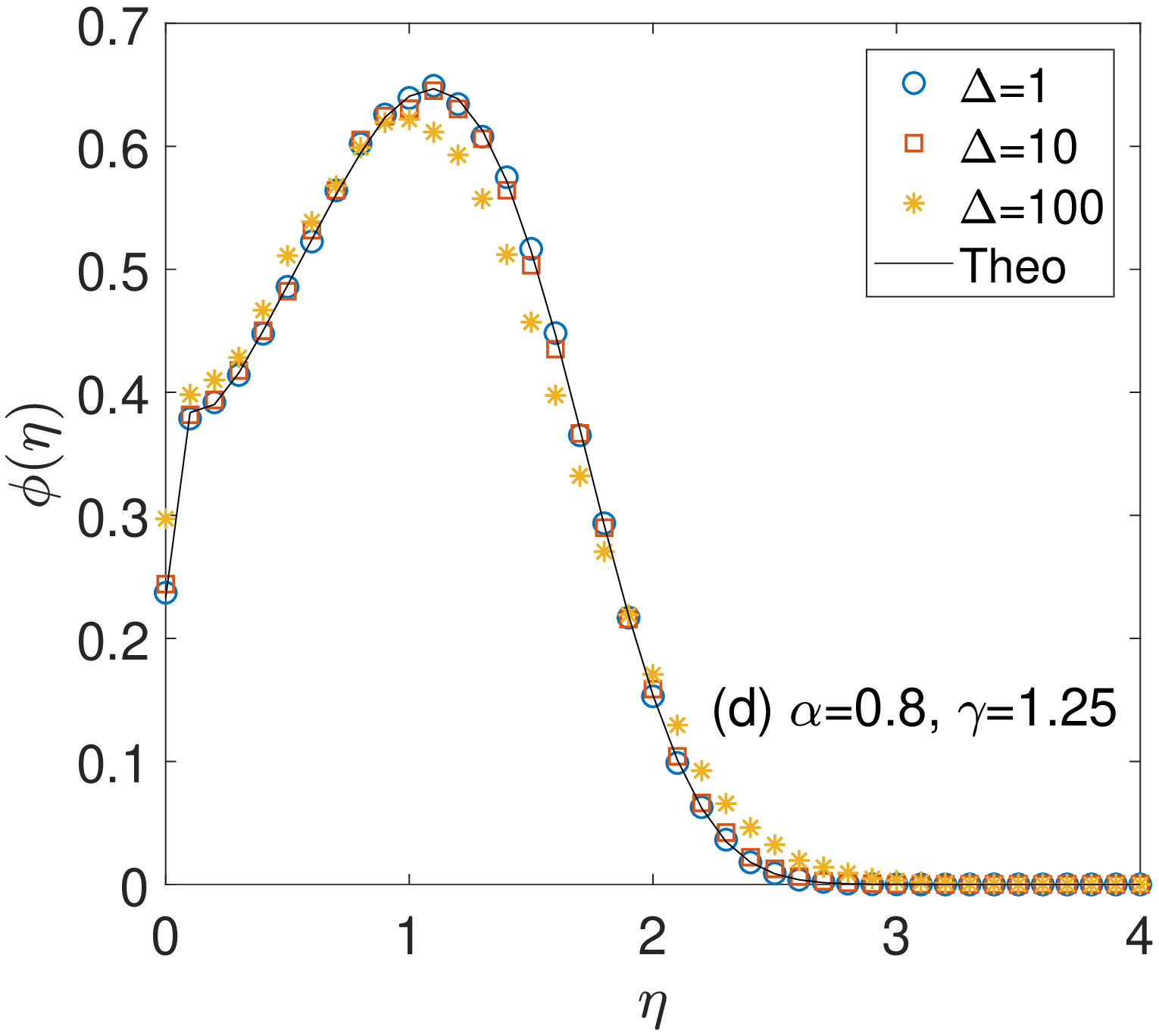}}
  \centerline{}
\end{minipage}
\caption{(Color online) Amplitude scatter PDF $\phi(\eta)$ for four kinds of parameters ($\alpha,\gamma$). The markers (circle, square, star) denote the simulations with different lag time $\Delta$ ($=1,10,100$), respectively. The solid lines are obtained from the theoretical result in Eq. \eqref{PDFTAMSD2} with the detailed algorithm presenting in Appendix \ref{App0}.  Due to the condition $\Delta\ll T$, the circle markers ($\Delta=1$) are more consistent to the solid lines than the star markers ($\Delta=100$). The dimensionless variable $\eta$ in four cases all present a broad distribution, implying the nonergodic property. Other parameters: the measurement time is $T=10^5$, and the number of trajectories used for ensemble is $10^5$.}\label{fig3}
\end{figure}

Similar to the discussion of EB parameter, we also assume that the diffusivity is uncorrelated at different times when evaluating the PDF $\phi(\eta)$ of TAMSD of model Eq. \eqref{DDmodel}. From the discussion of EB parameter above, we find that the uncorrelated random diffusivity acts as a deterministic one when we perform the
time averaging. Therefore, according to the general expression of TAMSD in Eq. \eqref{TAMSD2}, it can be evaluated as
\begin{equation}\label{TAMSD3}
    \overline{\delta^2(\Delta)}\simeq \langle\overline{\delta^2(\Delta)}\rangle_D
    :=\frac{2\Delta}{T}\int_0^{s(T)} \langle D(s')\rangle ds',
\end{equation}
where the symbol $\langle\cdot\rangle_D$ denotes the ensemble average over diffusivity $D(s)$. Substituting Eq. \eqref{MeanD} into Eq. \eqref{TAMSD3} yields
\begin{equation}
  \overline{\delta^2(\Delta)}\simeq  \frac{2\Delta}{T}s^\gamma(T),
\end{equation}
where $s(T)$ is random and denotes the value of the inverse subordinator $s(t)$ at time $t=T$. Thus, the dimensionless random variable in Eq. \eqref{Def-eta} can be denoted as
\begin{equation}
  \eta\simeq \frac{s^\gamma(T)}{\langle s^\gamma(T)\rangle},
\end{equation}
the PDF of which is
\begin{equation}\label{PDFTAMSD}
  \phi(\eta)=\frac{\Gamma^{1/\alpha\gamma}(\alpha\gamma+1)}{\alpha\gamma\Gamma^{1/\alpha\gamma}(\gamma+1)\eta^{1+1/\alpha\gamma}}
  L_\alpha\left[\frac{\Gamma^{1/\alpha\gamma}(\alpha\gamma+1)}{\Gamma^{1/\alpha\gamma}(\gamma+1)\eta^{1/\alpha\gamma}}\right],
\end{equation}
with the detailed calculations presenting in Appendix \ref{App2}. Compared with the subdiffusive CTRW, the diffusivity makes the EAMSD change from $t^\alpha$ to $t^{\alpha\gamma}$ in Eq. \eqref{EAMSD-1}. However, the PDF $\phi(\eta)$ in Eq. \eqref{PDFTAMSD} cannot be obtained by simply replacing $\alpha$ by $\alpha\gamma$ in Eq. \eqref{MLD}.
The corresponding simulations of $\phi(\eta)$ are present in Fig. \ref{fig3} with four groups of parameters $(\alpha,\gamma)$. Consider the condition $\Delta\ll T$ for evaluating TAMSD, we take three different lag time $\Delta=1,10,100$ for comparison. The simulations with smaller $\Delta$ are more consistent to the theoretical lines in Fig. \ref{fig3}.

The Eq. \eqref{PDFTAMSD} returns to Eq. \eqref{MLD} when $\gamma=1$, which is consistent to the discussions about the EB parameter that the second case $\alpha\neq1,\gamma=1$ in Table \ref{table1} converges to the subdiffusive CTRW for sufficiently large $T$.
For the first case $\alpha=1,\gamma\neq1$, the PDF of $\eta$ is $\phi(\eta)=\delta(\eta-1)$, which cannot be directly obtained from Eq. \eqref{PDFTAMSD} since $\alpha$ should be less than $1$ in L\'{e}vy distribution. Instead, we can consider this case from the beginning of the calculations in Appendix \ref{App2}, i.e., the PDF $h(s,t)$ of the inverse subordinator $s(t)$. The characteristic function in Eq. \eqref{hslambda} at $\alpha=1$ is $e^{-s\lambda}$, implying that $h(s,t)=\delta(t-s)$. Thus, the PDF $\phi(\eta)$ is also a $\delta$-function based on the derivations in Appendix \ref{App2}.
As to the third case $\alpha\gamma=1$ ($\alpha<1,\gamma>1$), the corresponding PDF is
\begin{equation}\label{PDFTAMSD2}
  \phi(\eta)=\frac{1}{\Gamma(\gamma+1)\eta^{2}}
  L_\alpha\left[\frac{1}{\Gamma(\gamma+1)\eta}\right],
\end{equation}
not a $\delta$-function, which corresponds to the nonzero EB parameter in Eq. \eqref{EB2}. In Eq. \eqref{PDFTAMSD2}, the exponent $\alpha$ is the L\'{e}vy index controlling the shape of the distribution of the TAMSD, while $\gamma$ is related to the scaling of L\'{e}vy distribution determining the width of the distribution. Although both $\alpha$ and $\gamma$ affect the diffusion behavior of random diffusivity model Eq. \eqref{DDmodel}, they play different roles which can be shown concretely through the analyses on TAMSD (especially the EB parameter and the distribution of TAMSD), but cannot be clearly separated from the calculations of EAMSD in Eq. \eqref{EAMSD-2}.

\section{Summary}\label{Sec6}
Brownian yet non-Gaussian phenomena have been observed in a large range of complex systems. Instead of the constant diffusivity $D$ of classical Brownian motion, the random diffusivity $D(t)$ becomes the key of many existing models which explains this phenomena theoretically. On the other hand, the particle might undergo trapping events and get immobilised in complex media, which can be described by an inverse subdinator $s(t)$. Therefore, this paper considers such a Langevin system with random diffusivity $D(s)$ coupled with a subordinator in Eq. \eqref{DDmodel}. The main purpose is to investigate the ergodic property of this model by evaluating the EAMSD and TAMSD. To explore the detailed features of TAMSD, we also analytically derive  the EB parameter and the distribution $\phi(\eta)$ of TAMSD.

The EAMSD $\langle x^2(t)\rangle$ and ensemble-averaged TAMSD $\langle\overline{\delta^2(\Delta)}\rangle$ are obtained in Eqs. \eqref{EAMSDt3} and \eqref{EATAMSD2} for any kind of diffusivity $D(s)$. Both of them depend only on the mean diffusivity $\langle D(s)\rangle$. Whether the EAMSD presents normal diffusion or not, the ensemble-averaged TAMSD is normal for any $D(s)$. For further calculations, we assume the diffusivity behaves as $\langle D(s) \rangle\simeq \gamma s^{\gamma-1}$ and find the EAMSD scales as $t^{\alpha\gamma}$ in Eq. \eqref{EAMSD-2}.

One interesting thing is that the EAMSD is equal to the ensemble-averaged TAMSD for long time when $\alpha\gamma=1$, i.e., $\langle x^2(\Delta)\rangle\simeq\langle\overline{\delta^2(\Delta)}\rangle$, which seems present an ergodic behavior. To detect the nonergodic behavior of the random diffusivity model Eq. \eqref{DDmodel} for any $0<\alpha<1$ and observe more detailed information of TAMSD, we evaluate the EB parameter and the distribution $\phi(\eta)$ of TAMSD. The procedure contains the calculation of correlation function of diffusivity $\langle D(s_1)D(s_2)\rangle$. For convenience, we assume that the diffusivity is uncorrelated at different time, and thus, obtain the explicit expressions of the EB parameter in Eq. \eqref{EB1} and the PDF of dimensionless random variable $\phi(\eta)$ in Eq. \eqref{PDFTAMSD} for any $\alpha$ and $\gamma$. Neither the former tends to zero nor the latter converges to a $\delta$-function as the measurement time $T\rightarrow\infty$, which proves the nonergodic behavior of the random diffusivity model Eq. \eqref{DDmodel}.

The assumption of the uncorrelated diffusivity is reasonable to some extent. This kind of diffusivity has been discussed in Refs. \cite{CherstvyMetzler:2016}. For more general diffusivity correlated at different time, the correlation usually strengthens the nonergodicity of random diffusivity model, which has been studied explicitly in Ref. \cite{WangChen:2021}.
However, there is one kind of diffusivity, the square of Ornstein-Uhlenbeck process \cite{ChechkinSenoMetzlerSokolov:2017}, which is correlated at different time but it reaches the stationary for long time. In this case, the diffusivity acts like the uncorrelated one at large measurement time \cite{WangChen:2021}, and the Langevin system is also nonergodic.
In fact, by observing the expression of TAMSD in Eq. \eqref{TAMSD2}, the nonergodicity of random diffusivity model results from not only the random diffusivity $D(s)$, but also the inverse subordinator $s(t)$. As long as $\alpha<1$, the TAMSD is not reproductive and remains a random variable as the measurement time $T\rightarrow\infty$.

The TAMSD is linear with respect to the lag time $\Delta$ in our random diffusivity model as Eq. \eqref{TAMSD2} shows, which is different from the EAMSD. Beyond this paper, the deviation of TAMSD from EAMSD is common for anomalous diffusion processes and comes from many effects, such as the initial condition. The time average is in some sense an equilibrium measure, while the ensemble average is not. Therefore, the form does not depend on the initial condition while the latter does \cite{LeibovichBarkai:2013,WangChenDeng:2019-2}. However, if the system starts from equilibrium initial conditions, the EAMSD and TAMSD would behave similarly with respect to lag time \cite{KlafterZumofen:1993,AkimotoCherstvyMetzler:2018,WangChenDeng:2019,HidalgoBarkaiBurov:2021}.
More connections between the random diffusivity model and other anomalous diffusion processes will be discussed in the future.

\section*{Acknowledgments}
This work was supported by the National Natural Science
Foundation of China under Grant No. 12105145, the Natural Science Foundation of Jiangsu Province under Grant No. BK20210325, and
the Fundamental Research Funds for the Central Universities under Grants No. lzujbky-2020-it02.

\appendix
\section{Simulation algorithms}\label{App0}
When generate the trajectories of the model Eq. \eqref{DDmodel}, we assume $D(s)$ obeys the exponential distribution with its mean in Eq. \eqref{MeanD}.
Since the theoretical results of the EB parameters in Eq. \eqref{EB1} and the PDF of TAMSD in Eq. \eqref{PDFTAMSD} are obtained in the condition that the diffusivity $D(s)$ is uncorrelated at different times, we generate $D(s_i)$ independently at different time nodes $s_i,~i=1,\cdots,N$.

As the random diffusivity model Eq. \eqref{DDmodel} contains two kinds of time variables, physical time $t$ and operational time $s$, we need to establish two sets of time lattices to express the subordinator $t(s)$ and the inverse subordinator $s(t)$.
Based on the first Langevin equation in model Eq. \eqref{DDmodel}, we can generate the trajectories of the original process $x(s)$. Then combining it with the trajectories of the inverse subordinator yields the trajectories of the subordinated process $x(s(t_i))$.
The explicit numerical algorithm of generating an inverse subordinator and a subordinated process for different Langevin systems can be found in Refs. \cite{MagdziarzWeronWeron:2007,GajdaMagdziarz:2010,WangChenDeng:2019,ChenDeng:2021}.

To generate the theoretical lines in Fig. \ref{fig3}, we first generate a random variable $X$ obeying the L\'{e}vy distribution $L_\alpha(x)$ with $\alpha<1$ based on the algorithm in Ref. \cite{KlafterSokolov:2011}. Then taking
\begin{equation}
  Y=\frac{\Gamma(\alpha\gamma+1)}{\Gamma(\gamma+1)}X^{-\alpha\gamma},
\end{equation}
the random variable $Y$ obeys the distribution in Eq. \eqref{PDFTAMSD2}, i.e.,
\begin{equation}
  p_Y(y)=\frac{1}{\Gamma(\gamma+1)y^2}
  L_\alpha\left[\frac{1}{\Gamma(\gamma+1)y}\right].
\end{equation}
By using the large samples of random variable $Y$, we can make a bar chart to depict the distribution of $Y$, which is the theoretical lines in Fig. \ref{fig3}.

\section{EB parameter in Eq. \eqref{EB1}}\label{App1}
Since the diffusivity is uncorrelated at different time, substituting Eq. \eqref{DDcorr} into the $I(T)$ in Eq. \eqref{IT}, we obtain
\begin{equation}
  \begin{split}
    I(T)&=\int_0^\infty \int_0^s\int_0^s \langle D(s'_1)\rangle \langle D(s'_2)\rangle ds'_1ds'_2 h(s,T)ds  \\
    &\simeq \int_0^\infty s^{2\gamma} h(s,T)ds,
  \end{split}
\end{equation}
where we do not consider the case $s'_1=s'_2$ in the first line since the corresponding integral domain can be omitted in the double integral.
Then by using the expression of the PDF $h(s,T)$ of inverse subordinator in Laplace domain as Eq. \eqref{hslambda} shows,
we have
\begin{equation}
  \begin{split}
    I(\lambda)&\simeq \int_0^\infty s^{2\gamma} h(s,\lambda)ds \\
    &\simeq \Gamma(2\gamma+1)\lambda^{-1-2\alpha\gamma}.
  \end{split}
\end{equation}
Performing the inverse Laplace transform gives
\begin{equation}
  I(T)\simeq \frac{\Gamma(2\gamma+1)}{\Gamma(2\alpha\gamma+1)}T^{2\alpha\gamma}.
\end{equation}
Substituting the $I(T)$ above and $\langle x^2(T)\rangle$ in Eq. \eqref{EAMSD-2} into Eq. \eqref{Def-EB}, the EB parameter in Eq. \eqref{EB1} can be obtained.

\section{PDF of TAMSD in Eq. \eqref{PDFTAMSD}}\label{App2}
The PDF of the inverse subordinator $s(T)$ can be expressed as \cite{BauleFriedrich:2005}
\begin{equation}\label{B1}
  h(s,T)=\frac{1}{\alpha}\frac{T}{s^{1+1/\alpha}}L_\alpha\left(\frac{T}{s^{1/\alpha}}\right).
\end{equation}
According to the quantitative relation between $s^\gamma(T)$ and $s(T)$, the PDF $h_\gamma(s,t)$ of $s^\gamma(T)$ can be expressed through the one of $s(T)$, i.e.,
\begin{equation}\label{B2}
\begin{split}
  h_\gamma(s,T)&=\frac{1}{\gamma}s^{-1+1/\gamma}h(s^{1/\gamma},T) \\
  &=\frac{1}{\alpha\gamma}\frac{T}{s^{1+1/\alpha\gamma}}L_\alpha\left(\frac{T}{s^{1/\alpha\gamma}}\right).
\end{split}
\end{equation}
Then we calculate the mean of $s(T)$ by use of the PDF $h(s,T)$ in Laplace domain ($T\rightarrow\lambda$) in Eq. \eqref{hslambda}, which in detail, is
\begin{equation}\label{B3}
  \begin{split}
    \mathcal{L}\{\langle s^\gamma(T)\rangle\}
    &= \int_0^\infty s^\gamma h(s,\lambda)ds \\
    &=\Gamma(\gamma+1)\lambda^{-1-\alpha\gamma}.
  \end{split}
\end{equation}
Performing the inverse Laplace transform gives
\begin{equation}\label{B4}
  \langle s^\gamma(T)\rangle =\frac{\Gamma(\gamma+1)}{\Gamma(\alpha\gamma+1)}T^{\alpha\gamma}.
\end{equation}
Then combining Eqs. \eqref{B2} and \eqref{B4}, we can obtain the distribution of random ${s^\gamma(T)}/{\langle s^\gamma(T)\rangle}$, i.e., the distribution of $\eta$ in Eq. \eqref{PDFTAMSD}.


\bibliography{ReferenceCW}

\end{document}